\documentclass[aps,prb,twocolumn,superscriptaddress]{revtex4-2} 
\usepackage{amsmath}
\usepackage{amssymb}
\usepackage{graphicx}
\usepackage{xcolor}
\usepackage{bm}
\usepackage{booktabs}
\usepackage{float}
\usepackage{hyperref}
\hypersetup{colorlinks=true}
\usepackage{braket}
\usepackage[normalem]{ulem} %provides \sout command.
\usepackage{siunitx}    % For SI units 
\newcommand{\tx}[1]{\text{#1}}
\begin{document}

\title{Probing chaos and thermalization through out-of-time-ordered correlators in random field spin chains}
%\title{Probing Chaos and Thermalization: OTOC in Disordered Spin Chains}

\author{C Jisha }
\email{jisha.2021rph03@mnnit.ac.in}
\author{Shivam Mishra}
\email{shivam.2023rph01@mnnit.ac.in}
\author{Ravi Prakash}
\email{ravi.prakash@mnnit.ac.in}
	
\affiliation{Department of Physics, Motilal Nehru National Institute of Technology Allahabad - Prayagraj-211004, India}

\begin{abstract}
Out-of-time-ordered correlators (OTOCs) have emerged as a diagnostic of information scrambling and quantum chaos in many-body systems. We investigate the imprints of chaos in the dynamics of OTOCs in the Heisenberg spin-$1/2$ chain with random fields. The system is parameterized to exhibit a crossover from integrable to chaotic dynamics. We demonstrate numerically that the approach to saturation of the OTOC can distinguish between integrable and chaotic regimes, with a power-law $(1/t)$ relaxation for integrable systems and a higher-degree power-law decay $(1/t^\alpha; \alpha \ge 1)$ followed by an exponential relaxation for the chaotic regime. We further show that long-range spectral statistics, such as the number variance, are more effective in characterizing quantum chaos in the regime near saturation of OTOC. We also demonstrate that the relaxation and initial scrambling regimes exhibit distinct and universal features, with the former being sensitive and the latter being robust against different realizations of random-fields. The long-time saturation of OTOC also fluctuates with different realizations,  and its exact expression is derived through the Eigenstate Thermalization Hypothesis.

\end{abstract}
	
\maketitle

\section{Introduction}
Chaos in quantum systems has been a subject of interest for a long time. Unlike classically chaotic systems, closed quantum systems do not depart exponentially for any initial perturbation due to unitary dynamics. The field of quantum chaos gained importance after the introduction of Random Matrix Theory (RMT) \cite{oxford_rmt, mehta2004random, stockmann, Porter, Brody-1981} and the BGS conjecture \cite{BGS_PRL_1984, BGS_1984}, which states that spectra of quantum chaotic and non-integrable systems are similar to random matrices satisfying similar symmetry requirements.
Non-integrability in quantum systems arising from interactions among their several degrees of freedom manifests as strong correlations in the eigen-energy spectrum. The chaotic spectra have strong non-degeneracies and exhibit level repulsion. Integrable systems, on the other hand, lack such correlations, and their spectral statistics resemble those of an uncorrelated sequence of random numbers. These fluctuation statistics can be characterized through statistical measures such as the nearest neighbor spacing distribution (NNSD), number variance statistic, spectral rigidity, etc.

The fluctuation statistics of quantum chaotic systems are universal, subjected to invariance of systems under time reversal and rotation, and agree perfectly with those of random matrix ensembles belonging to similar symmetries. The chaotic systems show the Wigner-Dyson statistics, characterized by the NNSD $P(s) = (\pi s/2) \exp(-\pi s^2/4)$, and the logarithmic growth of the number variance, $\Sigma^2(r) \propto \ln(r)$. In contrast, integrable systems exhibit Poisson statistics with $P(s) = \exp(-s)$ and $\Sigma^2(r) = r$ for NNSD and number variance, respectively. RMT has now become a standard reference to determine whether a quantum system is chaotic. However, RMT does not provide enough insight into the dynamics of states and chaos at early times. The fundamental problem of the exponential departure of trajectories in classical chaotic systems and its absence in the quantum regime remains unexplained through RMT.

Out-of-time-ordered correlators (OTOCs), that refers to any out-of-time-ordered product of observables, \textit{e.g.}, $\langle A(t_1) B(t_2) A(t_3) B(t_4)\ldots \rangle$, where $A$ and $B$ do not commute and the $t_j$ are not in order, are applied in several problems to study quantum chaos at early times \cite{Shen-2017, Slagle-2017, Cotler-2017, Campisi-2017, Hashimoto-2017, Fan-2017, Keyserlingk-2018, Nahum-2018, Rozenbaum-2017, Jalabert-2018, Arul-Baker-2018, Rakovszky-2018, rammensee2018many, Saraceno-2018, Chen-2018, Moudgalya-2018, Haehl-2017, Gharibyan-2018,Carlos-2019}.
The back-and-forth evolution in time is a necessary condition to study chaos. For all practical purposes, the OTOC is defined by the squared commutator in Eq.~(\ref{eq-otoc-commutator}), which contains out-of-time-ordered products. The growth of OTOC is governed by the system and choice of operators, but exhibits several generic features that depend on the symmetry and integrability of the system. 
Since the commutator reduces to the Poisson bracket in the classical limit, the OTOC can act as a bridge between quantum and corresponding classical dynamics.

\begin{figure}[!h]
	\includegraphics[width=0.6\linewidth]{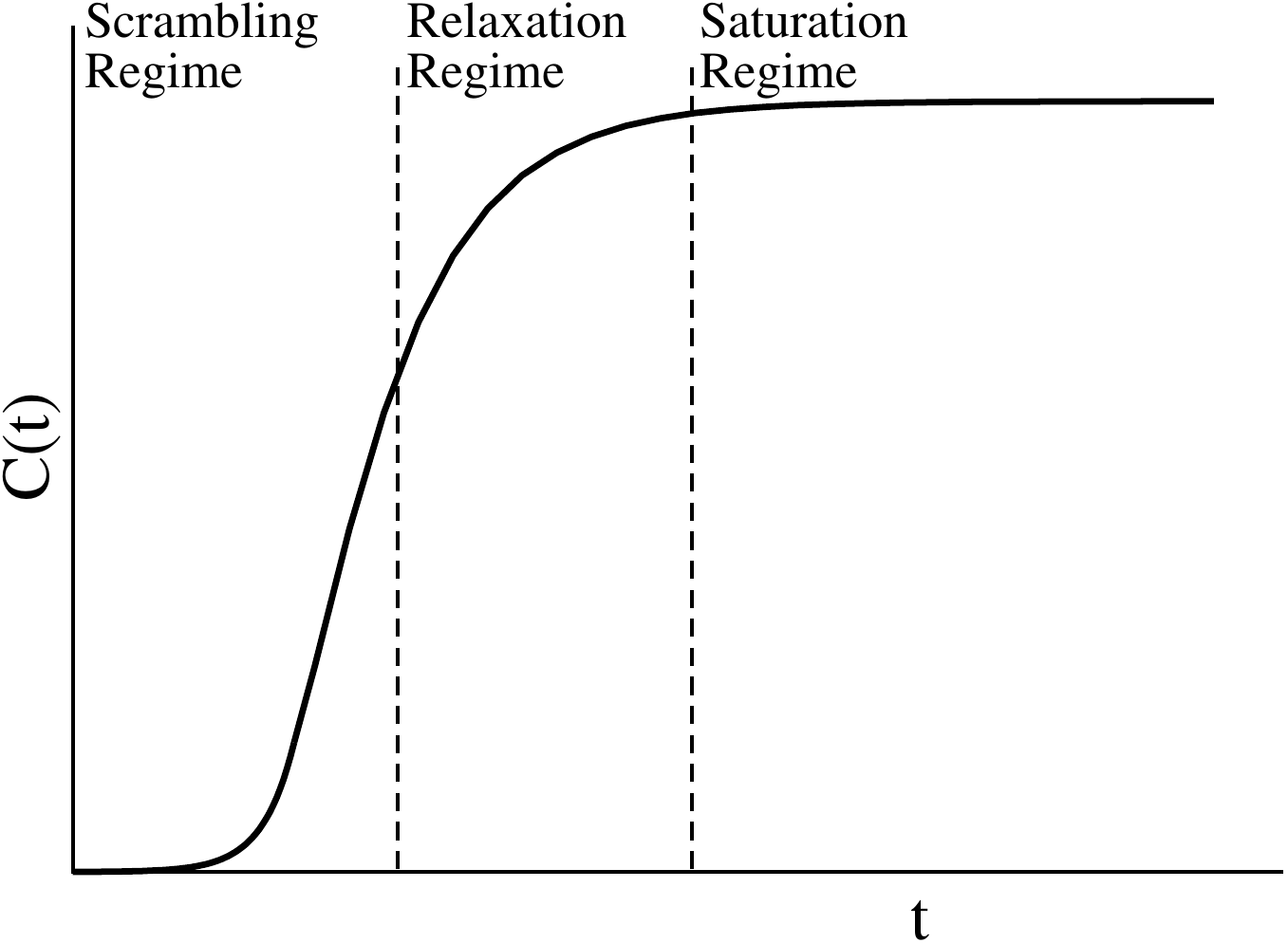}
	\caption{A schematic depicting the scrambling, relxation and saturation regime of OTOC.}
	\label{fig-otoc-schematic}
\end{figure}

For chaotic systems, the OTOC, in its initial scrambling regime (Fig.~(\ref{fig-otoc-schematic})), is expected to grow exponentially, with a rate twice the Lyapunov exponent \cite{maldacena2016bound}. Such an exponential growth has indeed been observed in several systems, including systems with a classical limit, \textit{e.g.}, quantum kicked rotor \cite{chaos2017}, coupled kicked rotors \cite{prakash-scrambling}, quantum maps \cite{lakshminarayan2019out}, and other systems \cite{chaos2017, Larkin-1969, Carlos-2019, maldacena2016bound}.
However, this behavior is not universal and is typically absent in many-body systems with only local interactions or without a classical limit. For example, spin chains \cite{riddell2019out, lee2019typical}, weakly chaotic systems \cite{kukuljan2017weak}, and Luttinger liquids \cite{dora2017out} exhibit a power-law growth of the OTOC both in chaotic and integrable regimes.

The OTOC eventually saturates at long times in finite-dimensional and bounded quantum systems due to the underlying unitary dynamics. In integrable systems, the OTOC typically exhibits large oscillations that arise from the presence of conserved quantities. In chaotic systems, it relaxes toward a steady saturation value, reflecting maximal scrambling. The saturation value can be estimated using RMT for chaotic systems \cite{prakash-scrambling, Balachandran-2021, PRB-block, lakshminarayan2019out, varikuti2024out}.

Note that, the OTOC continues to increase beyond the scrambling time before saturating. The corresponding intermediate regime between scrambling and saturation is referred to as the relaxation regime, as illustrated in Fig.~(\ref{fig-otoc-schematic}).  While the early-time exponential growth in semiclassical systems is relatively well understood, the reason for the subsequent increase in this regime is less clear and continues to be an active area of research. This growth rate is typically attributed to scrambling across the degrees of freedom driven by quantum interference effects \cite{rammensee2018many, PRXQuantum.5.010201}. Few-body systems with a classical limit and quantum maps, have been shown to exhibit power-law (exponential) relaxation for integrable (chaotic) systems \cite{prakash-scrambling, garcia2018chaos,rammensee2018many}. Such behavior can be analyzed in terms of Pollicot-Ruelle resonances of the corresponding classical dynamics \cite{venegeroles2007leading,fishman2002relaxation,garcia2018chaos}.
In contrast, many-body systems exhibit significantly more complex behavior. In \cite{Balachandran-2023}, the exponential relaxation is governed by the overlap of OTOC observables with the Hamiltonian, irrespective of the integrability or its absence. In \cite{Balachandran-2021}, the OTOC is shown to relax at most algebraically in presence of conserved quantities.  Floquet type kicked spin system has been shown to exhibit exponential relaxation for chaotic dynamics \cite{PRB-block}.

%\textcolor{red}{In The relaxation in such systems has been shown to depend on the overlap of OTOC operators with conserved quantities, rather than solely on integrability or its absence \cite{Balachandran-2023, Balachandran-2021}. Nevertheless, exponential relaxation has been observed in certain non-integrable Floquet spin systems \cite{PRB-block}}.

This article presents a comprehensive and numerical study of the OTOC in the time-independent Hamiltonian of Heisenberg spin chain with random fields, with the aim of identifying imprints of chaos in its dynamics, particularly in the relaxation regime. Our objective is to verify whether many-body systems lacking symmetries exhibit relaxation behavior similar to that observed in few-body systems \cite{prakash-scrambling, garcia2018chaos,rammensee2018many}. 
The system is parameterized to display a crossover from integrable to chaotic behavior. We demonstrate that the relaxation regime exhibits both power-law and exponential relaxation, and it can be helpful to distinguish between integrable and chaotic systems. We demonstrate that OTOC exhibits distinct and universal features in the scrambling and relaxation regimes, with the latter being sensitive to different realizations of random disorder fields and the former exhibiting no dependence on the growth rate. Similar behavior has also been reported in bipartite weakly coupled chaotic systems \cite{prakash-scrambling}, in the context of intra- and inter-subsystem scrambling. The long-time saturation of OTOC also fluctuates for different realizations. We derive an analytical expression for the saturation regime through the Eigenstate Thermalization Hypothesis (ETH).

This article is organized as follows. In section \ref{sec-model}, we describe the Heisenberg XXZ spin chain with non-integrability induced by random fields at each site. In section \ref{sec-spectral-properties}, we discuss the short-range and long-range spectral fluctuations and entanglement entropy. Section \ref{sec-otoc} presents our main results, where we discuss the OTOC and its universal behavior in the scrambling, relaxation, and saturation regimes. Finally, we summarize our findings in the section \ref{sec-summary}.

\section{Heisenberg spin chain} \label{sec-model}
The Hamiltonian for the spin-$1/2$ Heisenberg XXZ model is given by,
\begin{align}
\label{eq-Hxxz}
	H_\tx{I}  =  \sum_{i=1}^{L-1} \left[J \left(\sigma_i^x \sigma_{i+1}^x+ \sigma_i^y \sigma_{i+1}^y \right) + J_z \sigma_i^z \sigma_{i+1}^z \right] 
\end{align}
where $\sigma_i^{x,y,z}$ represents the $x,y,z$ component of the spin on the $i$-th site. The Hamiltonian $H_\tx{I}$ is integrable and can be solved analytically by the Bethe ansatz  \cite{betheansatz1998}. The system loses its symmetries and becomes non-integrable when magnetic fields are applied to one or more sites \cite{Kudo-2004}. We consider an ensemble of non-integrable systems with their Hamiltonian given by,% $H_\text{II} \equiv H_\text{II}(0)$, where
\begin{align}
\label{eq-Hnonintegrable}
H_\text{II} = H_\tx{I} + \sum_{i=1}^L h_{i}\sigma_i^x +  \sum_{i=1}^L g_i\sigma_i^z,
\end{align}
where $h_j = x_jk$ and $g_j = y_j k$ with $x$ and $y$ being uniform random variables in the interval $(-1,1)$. The ensemble consists of different realizations of uniform random values of $x$ and $y$. The field strength $k$ controls the  crossover between integrable and chaotic regimes.

For all numerical calculations, we consider the open boundary conditions for the spin chain. We take the flip-flop term, $J = 1 $, and the Ising term, $J_{z} = 0.85$.  The length of the chain is $L =12$. Since the number of random variables, $h_j$ and $g_j$, is only $2L = 24$, the set is not large enough to reflect a perfect random behavior. In order to ensure no correlation among random variables, we consider only those realizations of $h_j$ and $g_j$ that satisfy the following conditions: zero mean,\textit{ i.e.}, $\bar h = \left(\sum_j h_j\right)/L \approx 0$, $\bar g = \left(\sum_j g_j\right)/L  \approx 0$, and zero correlation factor, \textit{i.e.}, $C = \sum_j \left[\left(h_j - \bar h\right)\left(g_j -\bar g\right)\right]/\left(\sum_j (h_j - \bar h)^2  \sum_j (g_j - \bar g)^2\right)^{1/2} \approx 0$.
	
\section{Quantum Chaos Indicators}\label{sec-spectral-properties}
To diagnose the onset of quantum chaos in the ensemble, we study standard quantum chaos indicators based on spectral statistics and entanglement properties. As the disorder parameter $k$ is increased, the spin chain exhibits crossovers from integrable to chaotic and eventually to many-body localized (MBL) regimes.
To characterize spectral correlations at short and long ranges, we evaluate the NNSD and the number variance, respectively. The fluctuation statistics can be analyzed only after filtering out the smooth non-local variation of the average number density. The unfolding procedure rescales the spectra so that unfolded eigen-energies have unit number density everywhere. If the number density of spectra, averaged over the ensemble, is given by $dn/dE = \bar \rho(E)$, then unfolded eigen-energies for each spectrum ($\mathcal{E}$) can be obtained from the expression
\begin{align}
	\label{eq-unfolding}
	\mathcal E_j = \int_c^{E_j} \bar \rho (E) dE,
\end{align}
so that $dn/d\mathcal E = \bar \rho(\mathcal E) = 1$. Here, $c$ is an arbitrary constant that only shifts the complete spectrum and has no effect on spectral correlations.

\subsection{Nearest neighbor spacing distribution}
\begin{figure}[!h]
	\centering
	\includegraphics[width=0.8\linewidth]{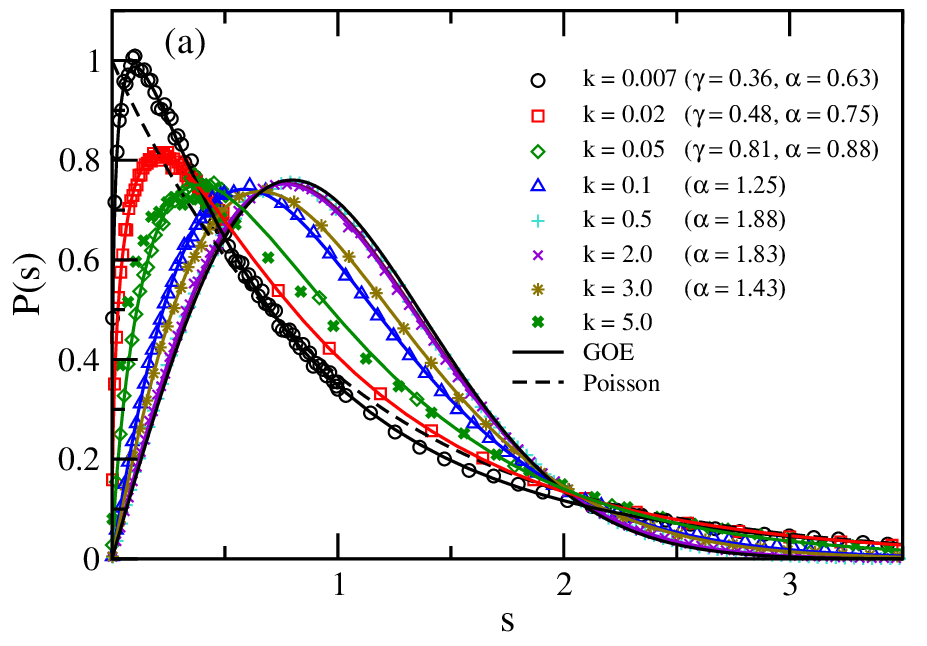}\\
%	\vspace{0.4cm}
	\includegraphics[width=0.45\linewidth]{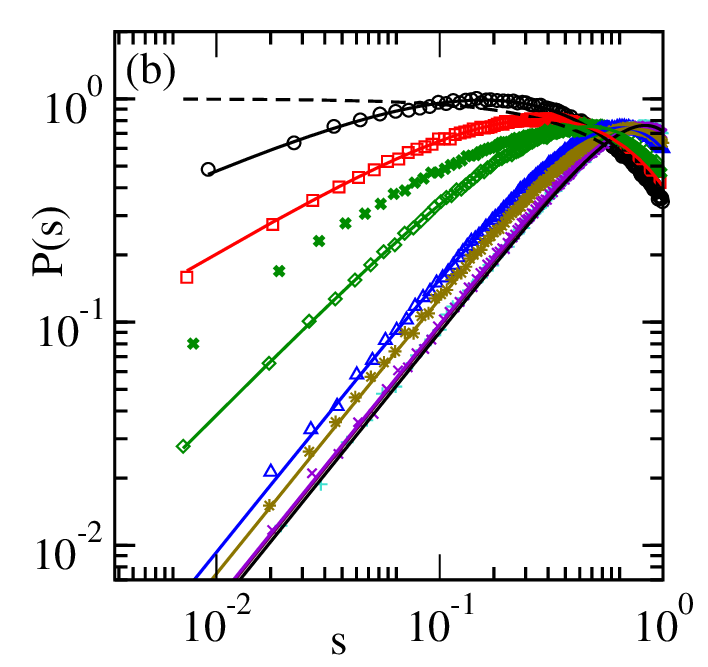}
	\includegraphics[width=0.51\linewidth]{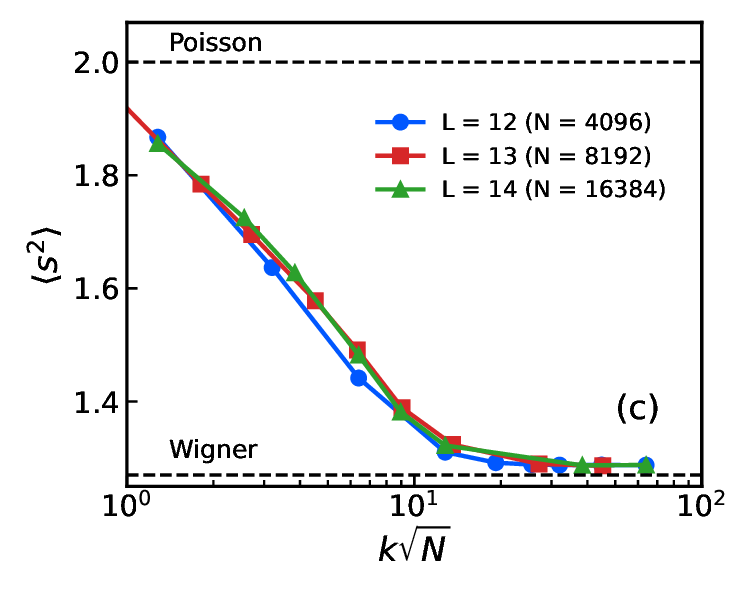}
	\caption{Spacing distributions for Hamiltonian $H_\text{II}$ for several values of field strength $k$ on (a) linear, and (b)log-log scale. The eigen-energies are unfolded using Eq.~(\ref{eq-unfolding}).
 Symbols represent the numerical data and solid lines represent the fitting curves for the spacing distribution, $P(s) = c_0 s^{\gamma} \exp(-c_1 s^\alpha)$. For $0.1 \le k \le 3$, $\gamma  =1$. (c) The crossover scales as $1/\sqrt{N}$ with system size. \label{fig-nnsd}}
\end{figure}

The short-range fluctuations can be analyzed using the NNSD of the unfolded spectra. The ensemble of systems $H_\text{II}$ is integrable when random fields $h_j$ and $g_j$ are zero or have very small values. These systems exhibit a transition from the Poisson to the Wigner distribution as the field strength $k$ is increased to break their symmetries and integrability.
The spacing distributions for different values of $k$ are shown in Fig.~(\ref{fig-nnsd}). The crossover scales as $1/\sqrt{N}$, with system size where $N = 2^L$ represents the dimension of matrices. We have used absolute values of $k$ in all our results. All these results are derived for $L = 12 (N = 4096)$, unless stated otherwise.
These systems are integrable for zero or very weak field strengths given by $0 \le k \le O(10^{-3})$, and show Poisson statistics for NNSD \cite{berrytabor}. As we further increase the parameter so that $O(10^{-3}) \le k \le O(10^{-2})$, onset of level repulsion is observed and nnsd follows $P(s) = c_0 s^{\gamma} \exp(-c_1 s^\alpha)$, where $0 < \alpha, \gamma < 1$. The constants $c_0$ and $c_1$ are evaluated from the normalization conditions, $\int P(s) ds = \int sP(s) ds = 1$.

For slightly higher values of $k$ in the range $0.1 \le k \le 0.5$, the NNSD follows the expression $P(s) = c_0 s \exp(-c_1 s^\alpha)$ very well, with $1 \le \alpha \le 2$ being the crossover parameter. As shown in Fig.~(\ref{fig-nnsd}), the onset of chaos is not uniform for all correlation lengths and appears very early for small spacings. The linear repulsion, \textit{i.e.}, $P(s) \propto s$ for small spacings, that is a key signature of a fully chaotic system, is observed for small field strengths, but at higher correlation lengths, the spacing distribution is still away from the Wigner-Dyson statistics and follows  $P(s) \propto \exp(-s^\alpha)$. The NNSD shows an excellent agreement with the Wigner distribution for $0.5 \le k \le 2$. The spacing distribution in different ranges of field strength is also summarized in Fig.~(\ref{fig-summary}).

If we further increase the field strength ($k > 2$), all the systems of the ensemble enter the MBL phase. The applied fields are now strong enough to localize the state to its initial condition. The chain behaves as if each site is uncorrelated from its neighboring site. The spectral statistics for MBL states, therefore, behave the same as those of integrable or uncorrelated systems.

\subsection{Number Variance Statistic}
\begin{figure}[!ht]
\includegraphics[width=0.9\linewidth]{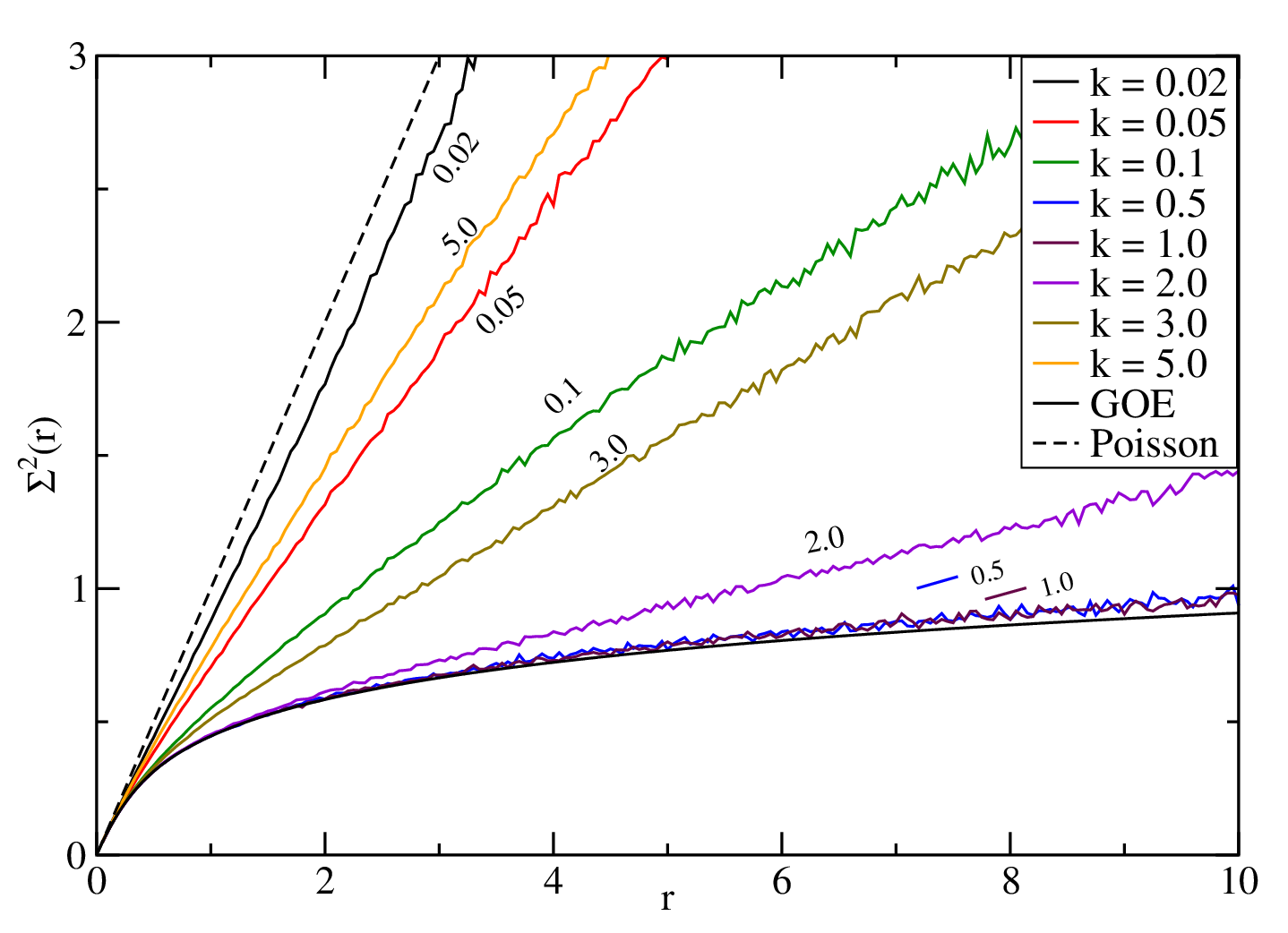}
\caption{Number variance statistic for different field strengths.
%	 An ensemble average is considered to suppress fluctuations due to finite-size Hilbert space. The number variance is consistent with the spacing distribution. For very small values of $k$, $0 \le k \le 0.05$ number variance is close to linear. For $0.5 \le k \le 1.0$ it is almost same as that of GOE. The intermediate values, $0.05 \le k \le 0.5$ consist of a mix behavior with small correlation lengths showing a logarithmic growth but linear at higher correlation lengths. MBL effects are observed for $k > 1$, where number variance starts shifting towards Poisson statistics.
}
\label{fig-nvar}
\end{figure}

The number variance measures the fluctuations in the count of unfolded eigen-energies in intervals of equal length, $r$. If a spectrum is divided into $m$ intervals of length $r$ and $n_1, n_2, \ldots, n_m$ represent numbers of eigen-energies in these intervals, then number variance is  defined as
\begin{align}
	\nonumber
	\Sigma^2(r) & = \langle n^2 \rangle - \langle n \rangle^2, \\
	\text{where}~\langle n^2 \rangle & = \frac{1}{m}\sum_{j = 1}^m n_j^2, ~\text{and}~ \langle n \rangle = \frac{1}{m} \sum_{j = 1}^m n_j = r.
\end{align}
Thus, number variance is useful to analyze the correlation among eigen-energies separated by a distance $r$. For integrable systems, due to the absence of correlations, the number variance increases linearly, $\Sigma^2(r) = r$, whereas for chaotic systems, due to strong spectral correlations, it is the same as that of Gaussian Orthogonal Ensemble (GOE) from RMT, which increases only logarithmically, $\Sigma^2(r) = 2\left(\ln(2\pi r) + \gamma + 1 -\pi^2/8\right)/\pi^2$ \cite{mehta2004random}.
The number variance for different values of field strengths, $k$, is plotted in Fig.~(\ref{fig-nvar}).

For small field strengths, $k \le 0.05$, the number variance is nearly linear, displaying the nearly integrable structure of the system. The transition from integrability to chaos takes place for $0.05 \le k \le 0.5$. Similar to what is observed in spacing distribution, number variance during the crossover shows logarithmic growth for small correlation lengths—a signature of strong chaos—but becomes linear at higher correlation lengths, exhibiting integrability at large correlation lengths. For $0.5 \le k \le 1.0$, the number variance shows an excellent agreement with GOE, indicating that the system is chaotic for short as well as moderately large correlation lengths. If we further increase $k$, the number variance shifts back to the integrable regime due to MBL.

\subsection{Entanglement Entropy}
\begin{figure}[!h]
	\includegraphics[width=0.9\linewidth]{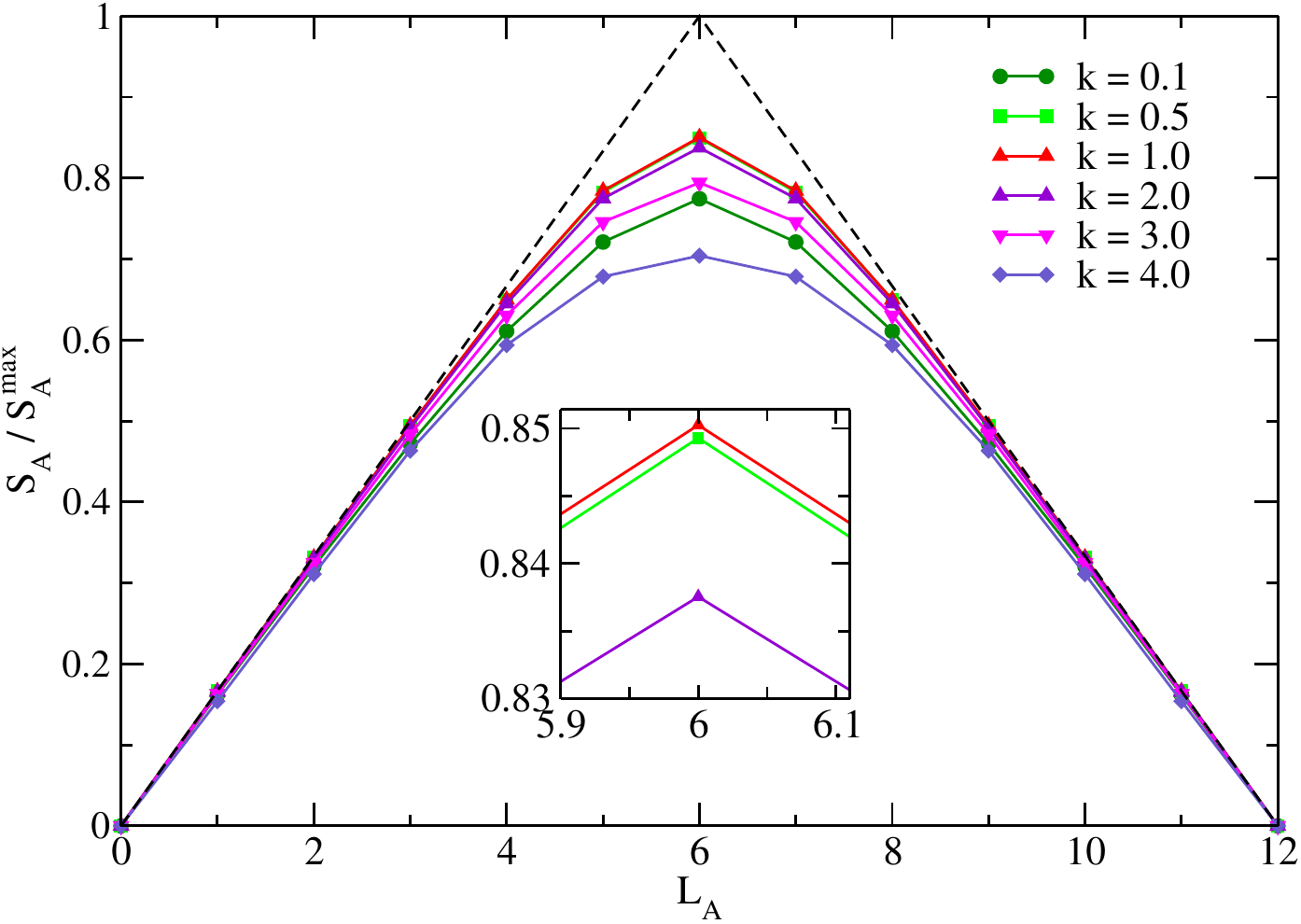}
	\caption{Average entanglement entropy of midspectrum eigenstates for the Hamiltonian $H_\text{II}$. The size of the subsystem $S_{A}$ is varied as $L_{A} = 0,1...,12$ for values of the crossover parameter $0.1\leq k \leq 4$. The dashed line represents the maximum entanglement entropy $S^{max}_{A}$.}
	\label{fig-page-curve}
\end{figure}

The thermalization of the spin chain arising from non-integrability, as well as the localization that occurs at sufficiently strong on-site fields, can be analyzed through the entanglement among the spins \cite{alet2018many,santos2004entanglement,sahu2025information}.
In the thermal (ergodic) phase, highly excited eigenstates exhibit volume-law entanglement, where the entropy grows proportionally with the subsystem size and approaches the Page value expected for random pure states \cite{page1993average}. In contrast, the MBL phase obeys an area law even at finite energy density, with entanglement entropy remaining bounded and significantly smaller than the thermal value, signaling the breakdown of ergodicity. We consider a bipartition of the spin-chain, given by the Hamiltonian $H_\text{II}$, into two subsystems $A$ and $B$ consisting of $L_A$ and $L_B = L - L_A$ sites, respectively. The entanglement entropy between $A$ and $B$ for a pure state $\ket{n}$ or the corresponding density matrix $\rho = \ket{n} \bra{n}$ is given by
\begin{align}
	\label{eq-entanglement}
	%\nonumber
	S_A = -\text{Tr}(\rho_{A}\ln \rho_{A})
\end{align}
where $\rho_{A} = \text{Tr}_{B}(\rho)$ is the reduced density matrix of the subsystem $A$ obtained after taking the partial trace on the degree of freedom of subsystem $B$.
Note that $S_A = S_B$. We compute the average entanglement entropy of mid-spectrum eigenstates of  $H_\text{II}$. The entropy does not change for different realizations of random fields with the same field strength, $k$.  The Page curves for different values of $k$ are normalized by the maximal possible value of entanglement entropy, $S^{max}_{A} = (L/2) \ln 2 \approx 4.1589$, which serves as an upper bound.  For a weak disorder ($k \le 0.1$), the entanglement entropy grows slowly and remains much below the maximal value, indicating incomplete mixing of the subsystem degrees of freedom. As the field strength is increased, $0.5\leq k \leq 1$, the entropy approaches the Page value, $S \approxeq ln L_{A} - L_{A} / 2L_{B}$ (\textit{i.e.}, the typical volume-law value for random pure states), which is strictly smaller than $S^{max}_{A}$. This indicates a thermal phase governed by the ETH as shown in Fig.~(\ref{fig-page-curve}). Upon further increasing $k$ ($k > 1$), strong disorder suppresses ergodicity and drives the system towards the MBL phase. In this regime, the entanglement entropy is significantly reduced compared to the thermal phase, reflecting the breakdown of volume-law scaling and the emergence of localization.

\begin{figure*}[ht]
	\includegraphics[width=0.9\linewidth]{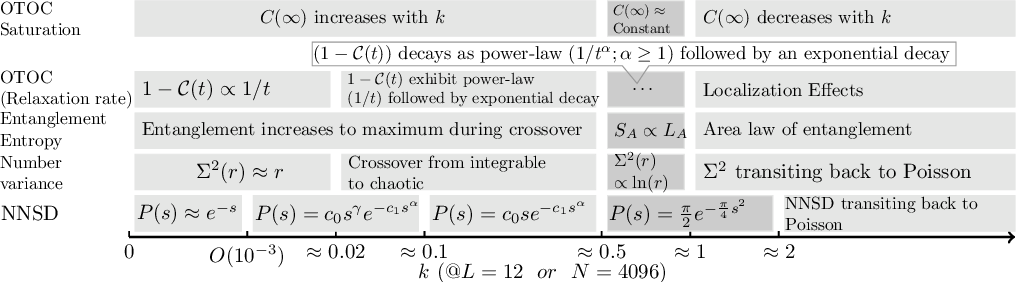}
	\caption{A schematic to summarize crossover from integrability to chaos in NNSD, number variance statistic, entanglement entropy, and OTOC. Dark shaded blocks correspond to agreement with RMT or completely thermalized cases.}
	\label{fig-summary}
\end{figure*}

To this end, we observe that the spacing distribution shows an excellent agreement with RMT over a wide range of field strengths $0.5 \le k \le 2$, whereas the number variance and entanglement entropy agree with RMT only in a narrower range $0.5 \le k \le 1$. See Fig.~(\ref{fig-summary}) for a quick comparison among NNSD, number variance, entanglement entropy, and OTOC. The corresponding imprints of chaos are expected to occur in OTOC at a time scale $t \approx \delta E/\hbar$, and therefore long-range correlations are more suitable to analyze chaos in the relaxation regime. We shall show ahead that the relaxation regime treats the ensemble as chaotic only for $0.5 \le k \le 1$.

\section{Out-of-time-ordered correlators} \label{sec-otoc}
Information scrambling in quantum many-body systems refers to the process by which local quantum information spreads across the entire system due to entanglement and interactions, effectively making the information inaccessible to local measurements. OTOCs have been in use for a long time to study information scrambling. The OTOC, $C(t)$ is defined for two observables $A$ and $B$ as,
\begin{align}
	\label{eq-otoc-commutator}
	C(t) & = -\langle[A(t),B(0)]^2\rangle,
	%& = -\text{Tr} [A(t),B(0)]^2 ~~(\text{for}~ T \rightarrow \infty)
\end{align}
where $\langle \cdot \rangle$ denotes the thermal statistical average, $\langle O \rangle = \text{Tr}[\exp(-\beta H) O]$, at temperature $\beta = 1/kT$. We consider the infinite temperature limit so that OTOC is given by $C(t) = -\text{Tr} [A(t),B(0)]^2$, where $A(t) = e^{iHt}~A~e^{-iHt}$ represents the operator at time $t$ in the Heisenberg representation. The operators $A$ and $B$ are chosen to span different sites. Initially, $C(t)$ is zero because of no overlap between the operators, and it increases when $A(t)$ starts spreading and overlaps with $B$.
We categorize the OTOC in three temporal regimes and examine the universality of its behavior in each of them (see Fig.~(\ref{fig-otoc-schematic})). The initial increase defines the scrambling regime, where operators spread across the available Hilbert space. At long times, the OTOC approaches a saturation value determined by the effective size of the Hilbert space, taking into account the symmetries of the system. Of particular interest in this work is the intermediate relaxation regime, where the OTOC evolves beyond the scrambling phase and relaxes toward its eventual saturation value. In the following sections, we analyze the OTOC behavior in these regimes in detail.

For the observables, we consider two types of operators to evaluate OTOC. In the first case, we choose operators that are localized on individual sites, \textit{e.g.},
\begin{align}
	A_1 = S_j^x~\text{and}~B_1 = S_k^z,
\end{align}
placed at sites $j$ and $k$ of the spin chain with $j \ne k$. In the numerical results discussed ahead, the operators $A_1$ and $B_1$ are placed at sites $3$ and $10$ respectively. 
We also consider block (non-local) operators that are initially spanned over  half of the spin chain \cite{PRB-block}.
\begin{align}
	\label{operator1}
	A_2 = \sum_{i=1}^{L/2} S_{x}^{i}, 
	~\text{and}~
	B_2 = \sum_{i=L/2 + 1}^L S_{z}^{i}
\end{align}

\subsection{Scrambling regime}
\begin{figure}[!ht]
	\includegraphics[width=\linewidth]{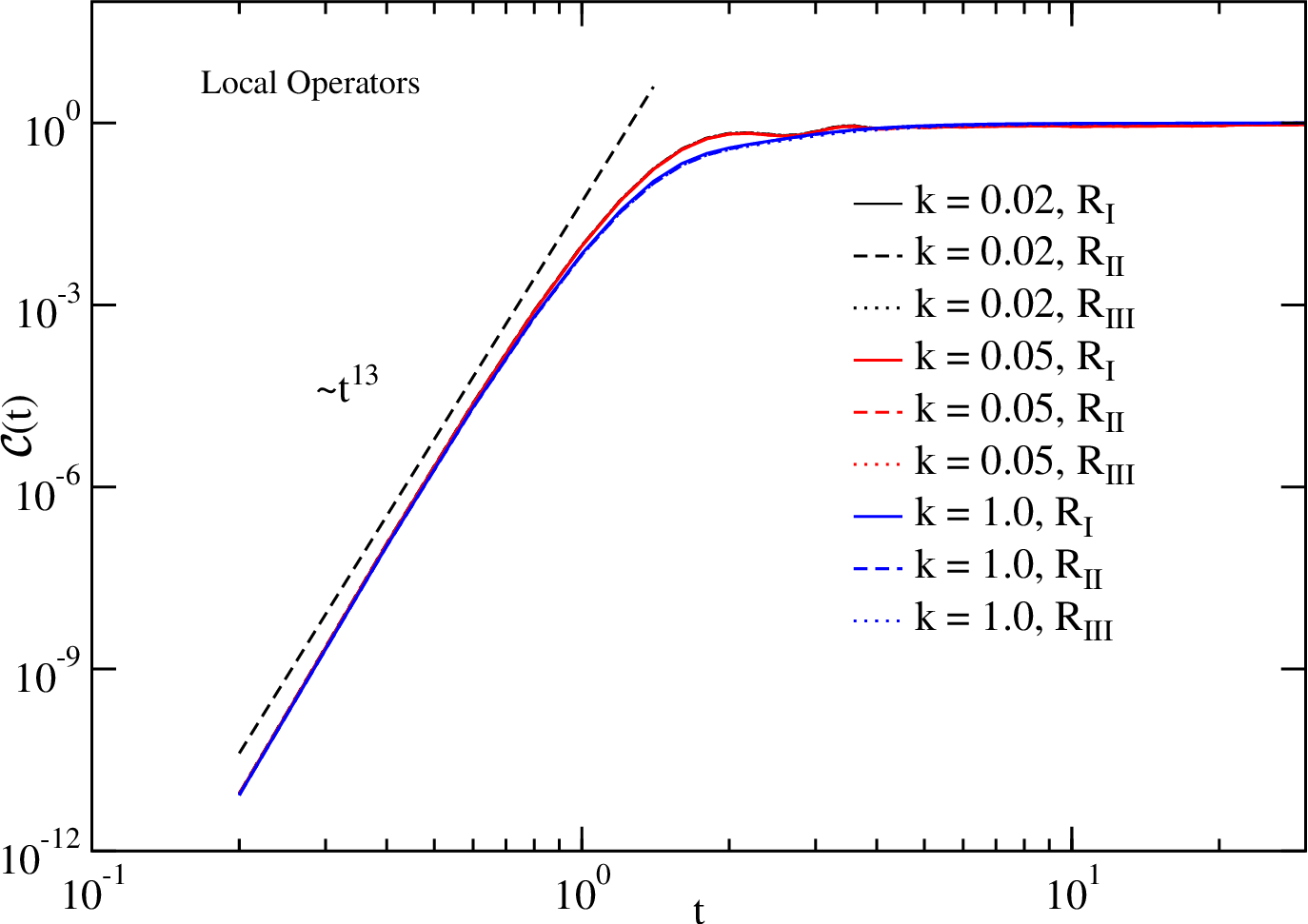}
	\caption{Normalized OTOC (on log-log scale) for local operators $A_1$ and $B_1$ evolving under the dynamics given by $H_\text{II}$. Three realizations, denoted by $R_j$ for random fields are shown for each value of $k$. %Inset shows same plot on linear scale.
	}
	\label{fig-otoc-k-all-local}
\end{figure}

\begin{figure}[!ht]
\includegraphics[width=\linewidth]{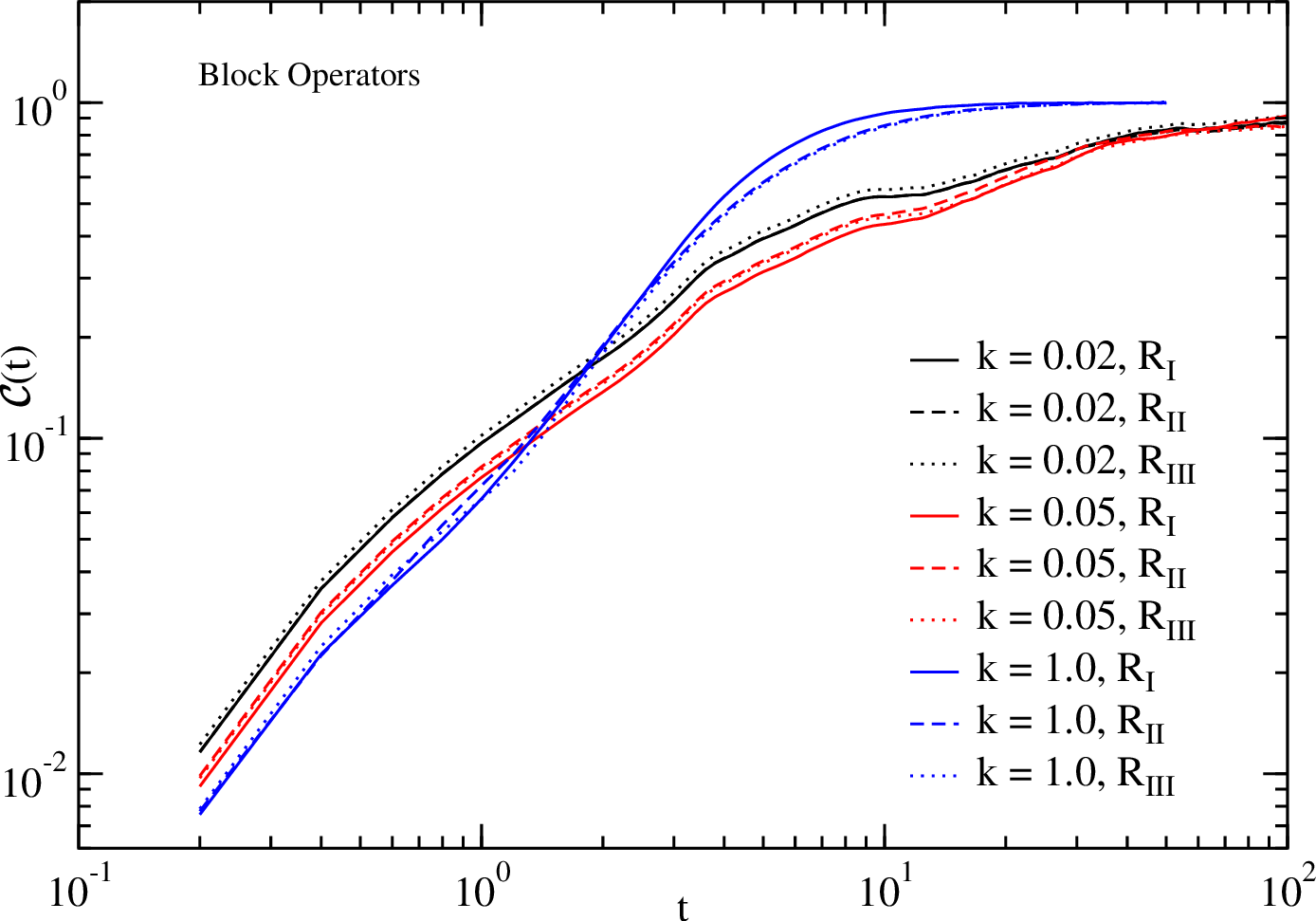}
	\caption{Normalized OTOC for block operators $A_2$ and $B_2$ with other conditions same as in Fig.~(\ref{fig-otoc-k-all-local}).}
	\label{fig-otoc-k-all-nonlocal}
\end{figure}

This regime typically involves the spreading of initially localized quantum information until the observables explore the entire available Hilbert space. Systems with a classical analog behave approximately in accordance with their corresponding classical dynamics in this regime. For classically chaotic systems, the growth of the OTOC is exponential, whereas integrable systems follow a power-law behavior. However, quantum many-body systems without any classical analog, such as spin-$1/2$ chains and other systems, do not differentiate between integrable and chaotic dynamics in this regime. Instead, both types of systems display power-law or algebraic growth of the OTOC \cite{Fortes-2019, Craps-2020}. In Figs.~(\ref{fig-otoc-k-all-local}, \ref{fig-otoc-k-all-nonlocal}), we plot the normalized OTOC, $\mathcal C(t) = C(t)/C(\infty)$ for local operators, $A_1$ and $B_1$, and block operators $A_2$ and $B_2$, for the dynamics generated by the Hamiltonian $H_\text{II}$. The saturation value, $C(\infty)$ is estimated by averaging the OTOC over multiple time steps at sufficiently long times ($t > O(1000)$).

We observe power-law growth of the OTOC for local operators, with a rate that depends on the separation between the operators \cite{lin2018out}. For the non-local operators, $A_2$ and $B_2$, the OTOC exhibits  algebraic growth instead of power-law behavior, possibly due to their non-local structure.

\textit{The growth rate depends on observables but not on random disorder fields:} As shown in Fig.~(\ref{fig-otoc-k-all-local}), the OTOC for local observables $A_1$ and $B_1$ remains the same for multiple realizations of random on-site fields $h_j$ and $g_j$. It is interesting to note that, for different disorder strengths ($k$) of the spin chain, the scrambling regime follows a power-law growth of the same order, failing to differentiate between chaos and integrability.

The OTOC for block operators is shown in Fig.~(\ref{fig-otoc-k-all-nonlocal}). The initial growth for a very-short time scale ($t = O(1)$), also known as the decay time, is power-law due to perturbative dynamics. The four-point correlator remains constant up to this time. After this time, the OTOC exhibits universal characteristics.
The block operators show a faster scrambling rate for $k = 1.0$ for which the system lies well within the chaotic regime as suggested by the quantum chaos indicators. The growth rate, for fix value of $k$, remains the same for different realizations. 

\subsection{Relaxation regime}
The OTOC, beyond the scrambling regime, gradually enters the relaxation regime, which occurs in the intermediate time window between scrambling and saturation. For systems possessing a classical limit, the relaxation regime is known to exhibit an exponential (power-law) relaxation for chaotic (integrable) systems \cite{prakash-scrambling, rammensee2018many, garcia2018chaos}. An important question, and a key motivation of the present work, is whether similar signatures of integrability and chaos persist in systems without a classical analog. 
We demonstrate that the relaxation regime carries signatures of the underlying spectral fluctuations of the system and is useful to distinguish between integrable and chaotic systems. We observe qualitatively different relaxation dynamics across the spectral crossover, with power-law $(1/t)$ decay in the integrable regime and an exponential decay in the chaotic regime. As summarized in Fig.~(\ref{fig-summary}), short and long-range chaos indicators transit from integrability to chaos at different field strengths. It implies that the onset of chaos occurs at different times in OTOC for different field strengths. While speaking of \textit{integrability} and \textit{chaos} in the context of relaxation regime of OTOC, we rely on the long-range fluctuation statistics, \textit{e.g.}, number variance statistic instead of NNSD, because energy correlations of order $\delta E > \hbar/\delta t$ are dominant in the early dynamics.

\begin{figure}[!ht]
\includegraphics[width=\linewidth]{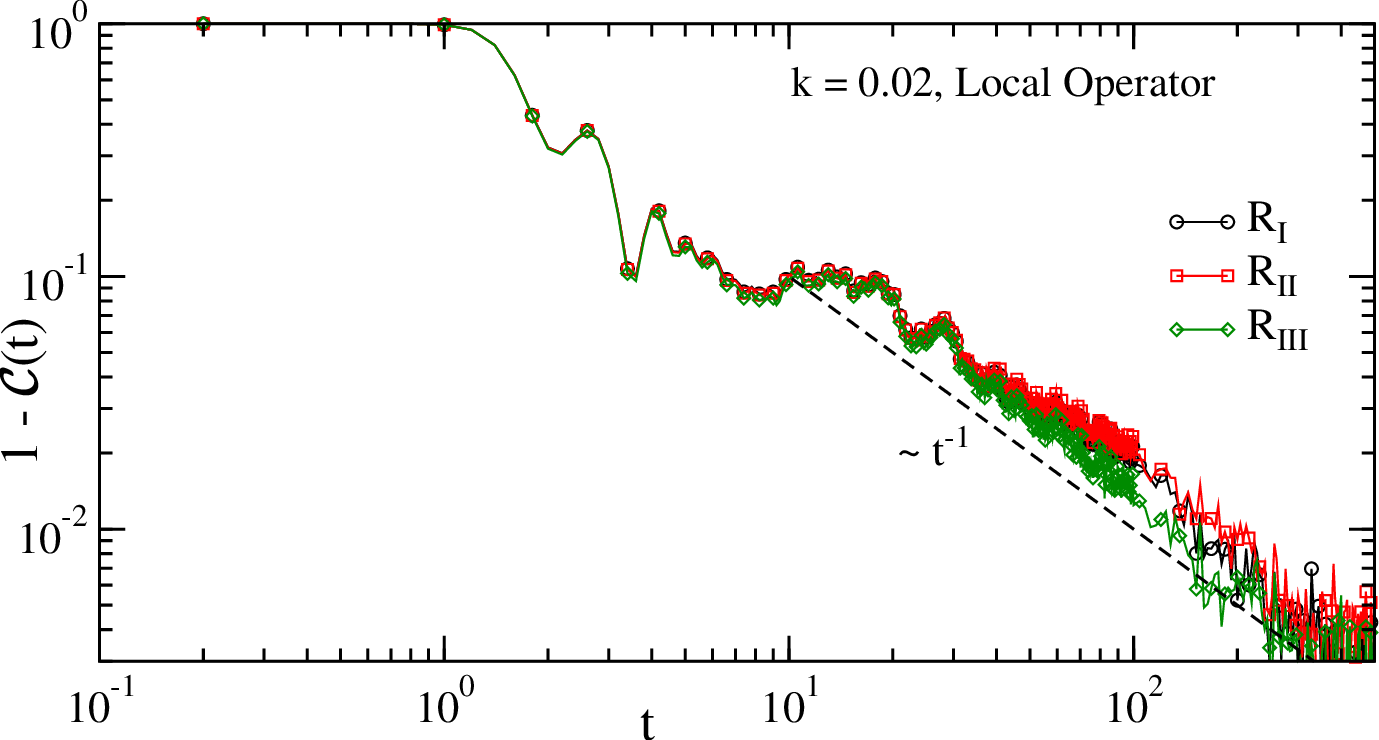}
\includegraphics[width=\linewidth]{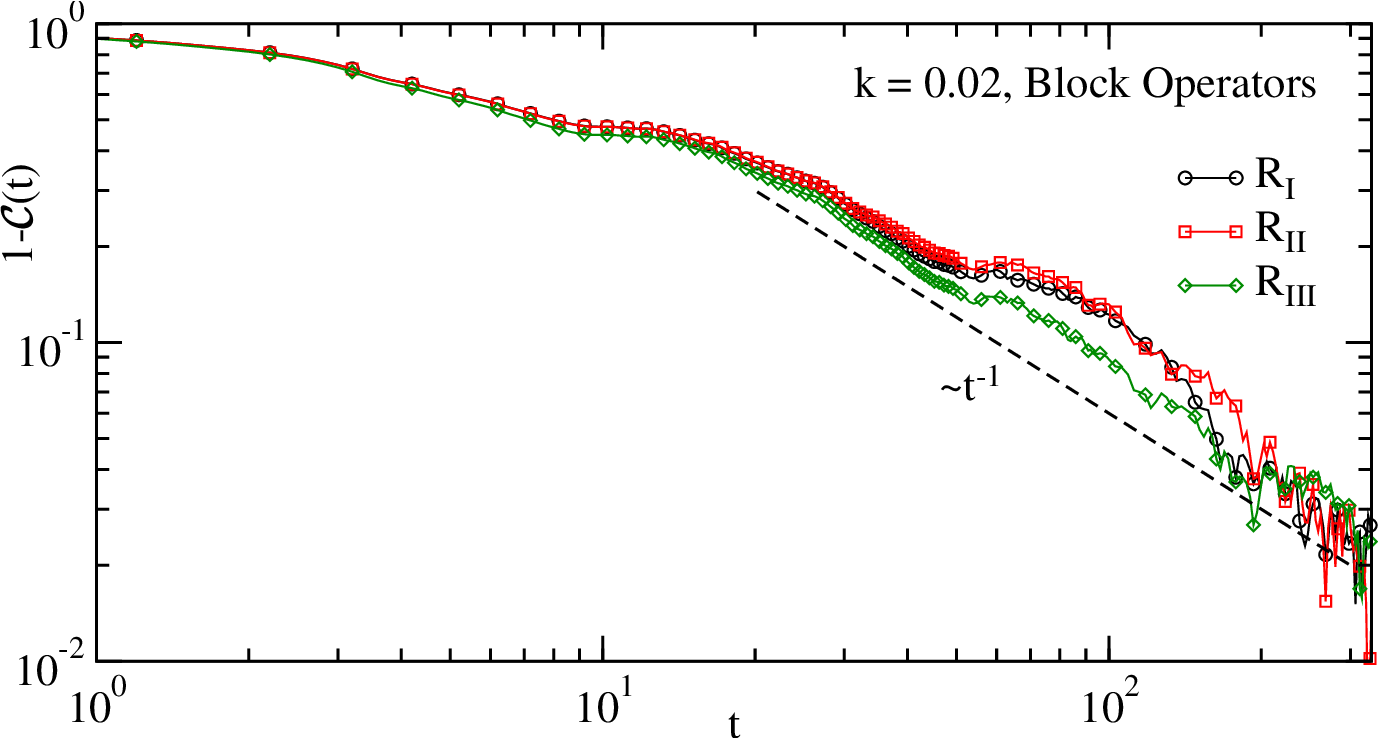}
	\caption{Relaxation dynamics of OTOC for field strength $k = 0.02$ with local ($A_1, B_1$) and block ($A_2, B_2$) operators. The plots, denoted by $R_\text{I}$ to $R_\text{III}$, correspond to different realizations of random fields on spin site. The OTOC relaxes as a power-law, $1/t$.}
	\label{fig-otoc-relax-local-nonlocal-k0.02}
\end{figure}
\begin{figure}[!ht]
	\includegraphics[width=\linewidth]{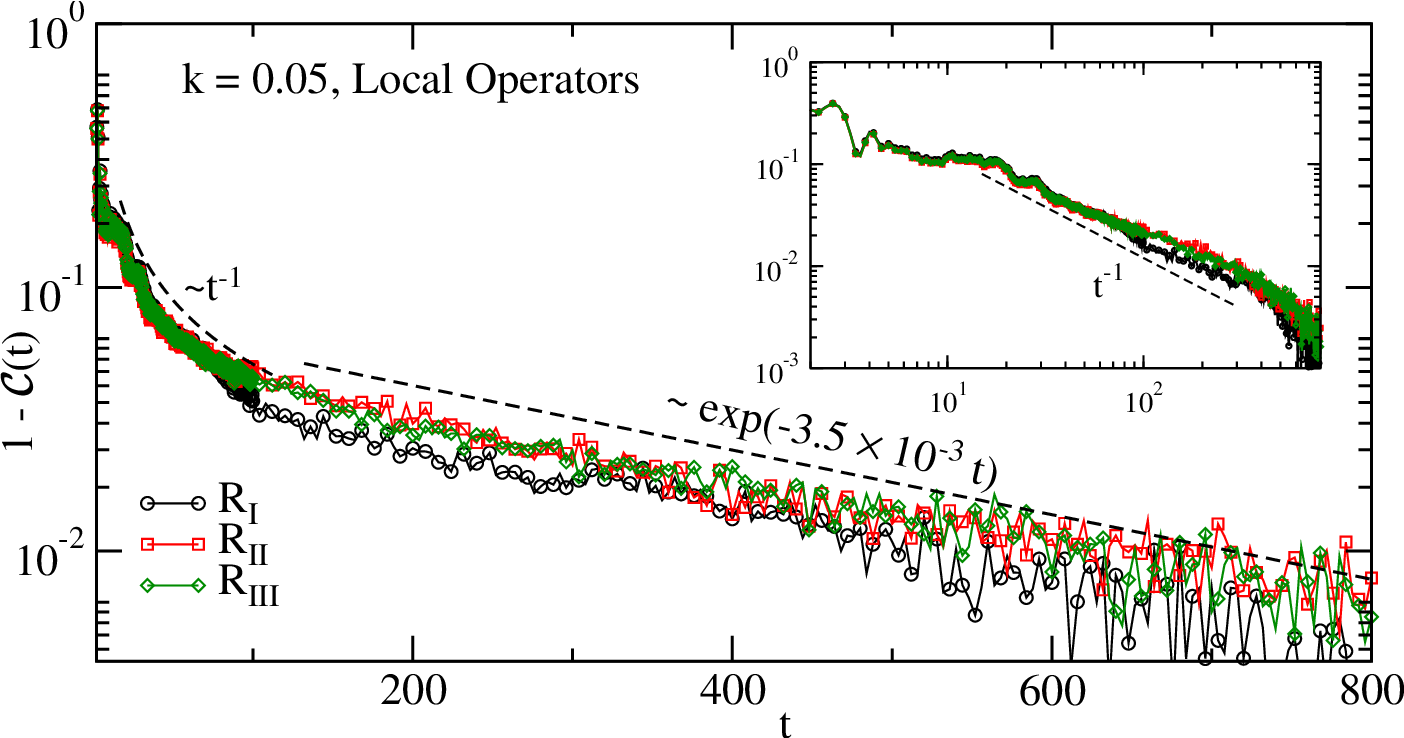}
	\includegraphics[width=\linewidth]{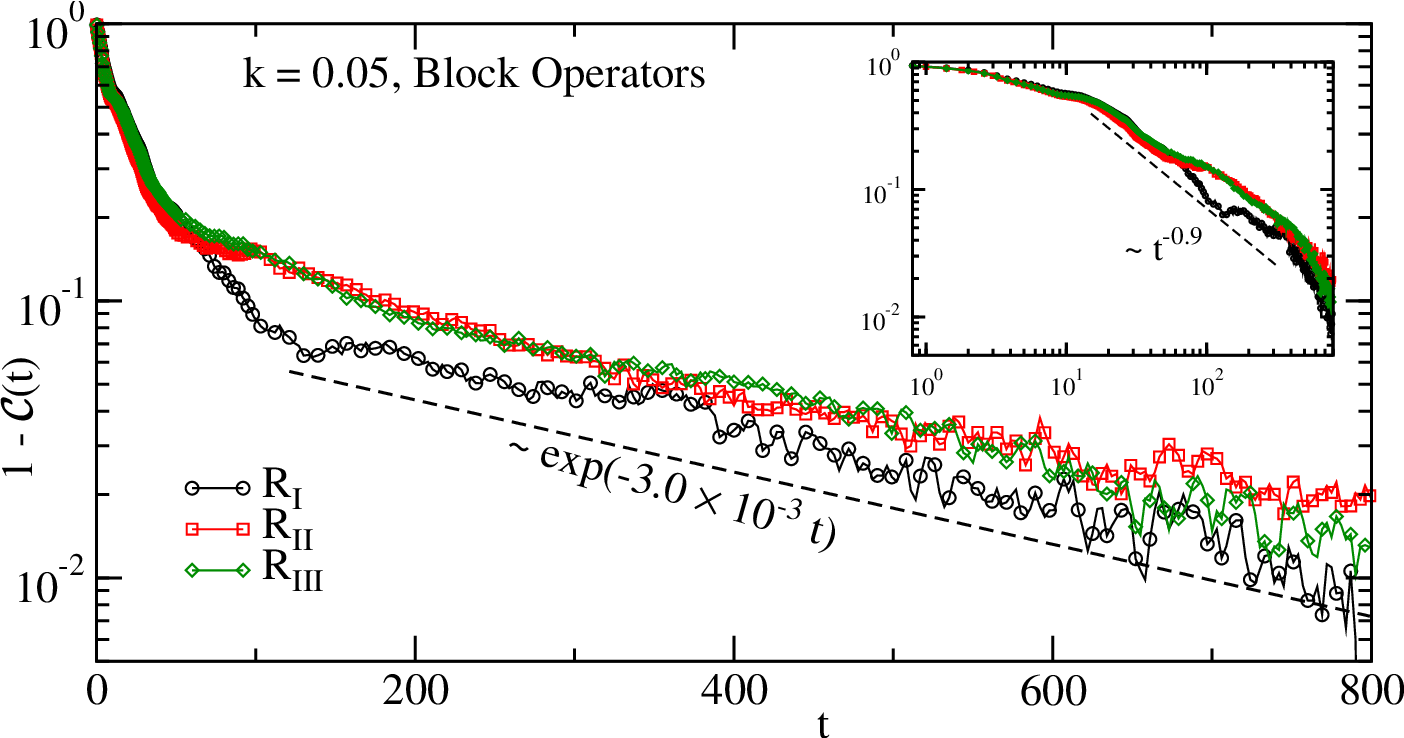}
	\caption{Relaxation dynamics of OTOC for field strengths $k = 0.05$ with local (top) as well as block (bottom) operators. The OTOC relaxes as power-law (exponentially) for local (block) operators but eventually follows an exponential decay for both type of operators. Since the decay rate is very small, it can also be considered a power-law relaxation. Inset show the same plot on log-log scale for comparison.}
\label{fig-otoc-relax-local-nonlocal-k0.05}
\end{figure}
\begin{figure}[!ht]
	\includegraphics[width=0.8\linewidth]{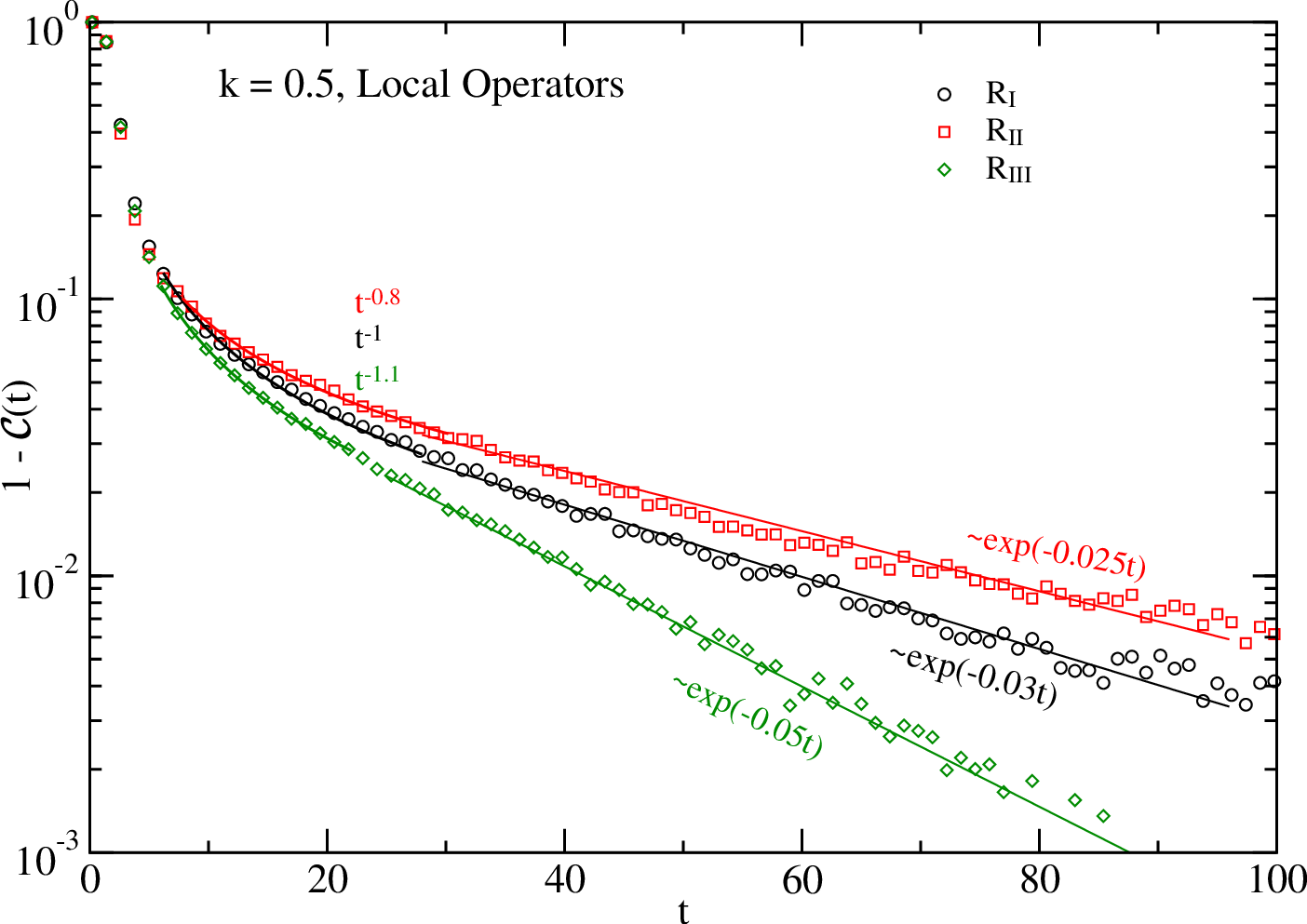}
	\includegraphics[width=0.8\linewidth]{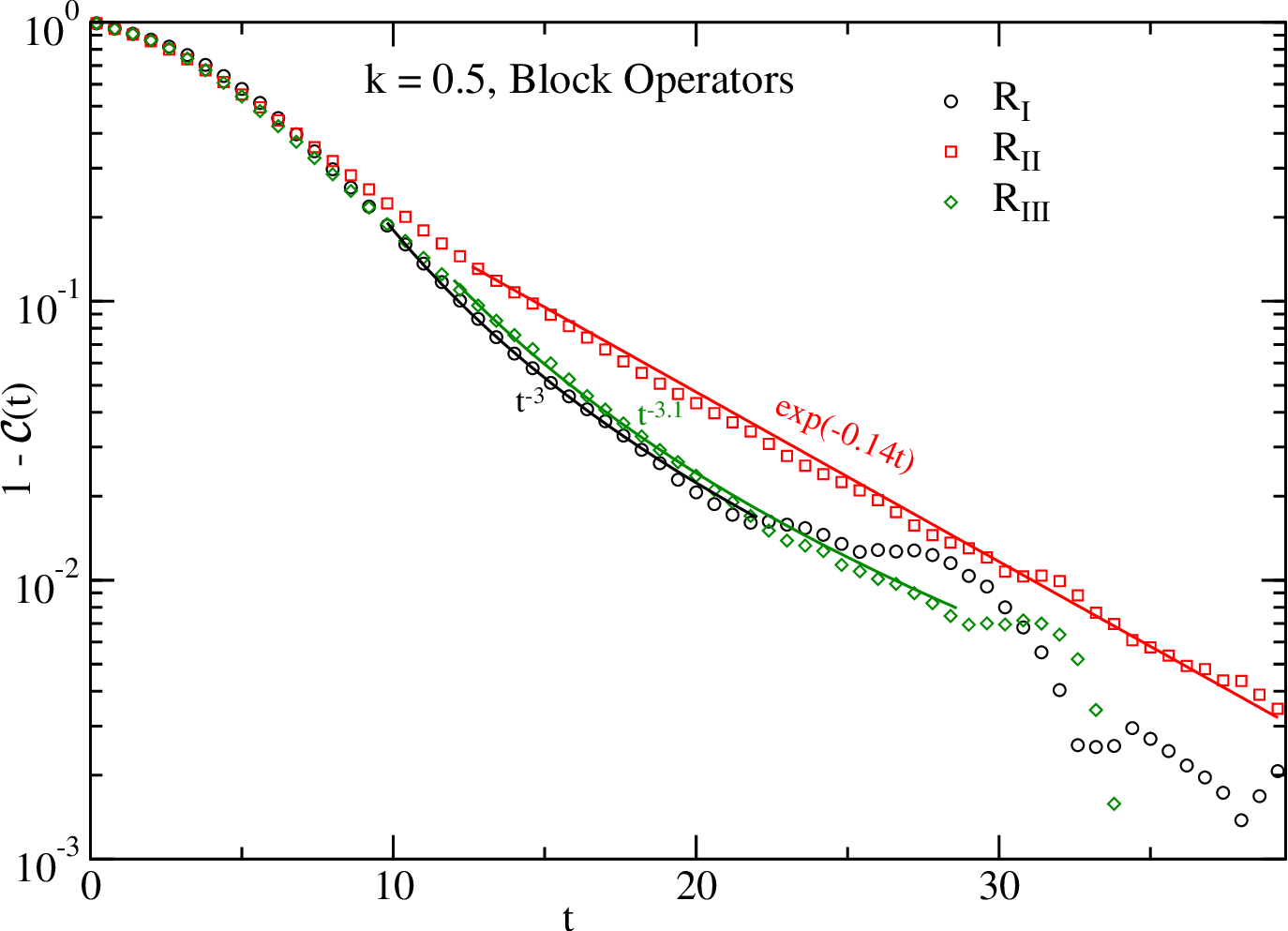}
	\caption{Relaxation dynamics of OTOC for field strength $k = 0.5$ with local and block operators.}
	\label{fig-otoc-relax-local-nonlocal-k0.5}
\end{figure}
\begin{figure}[!ht]
	\includegraphics[width=\linewidth]{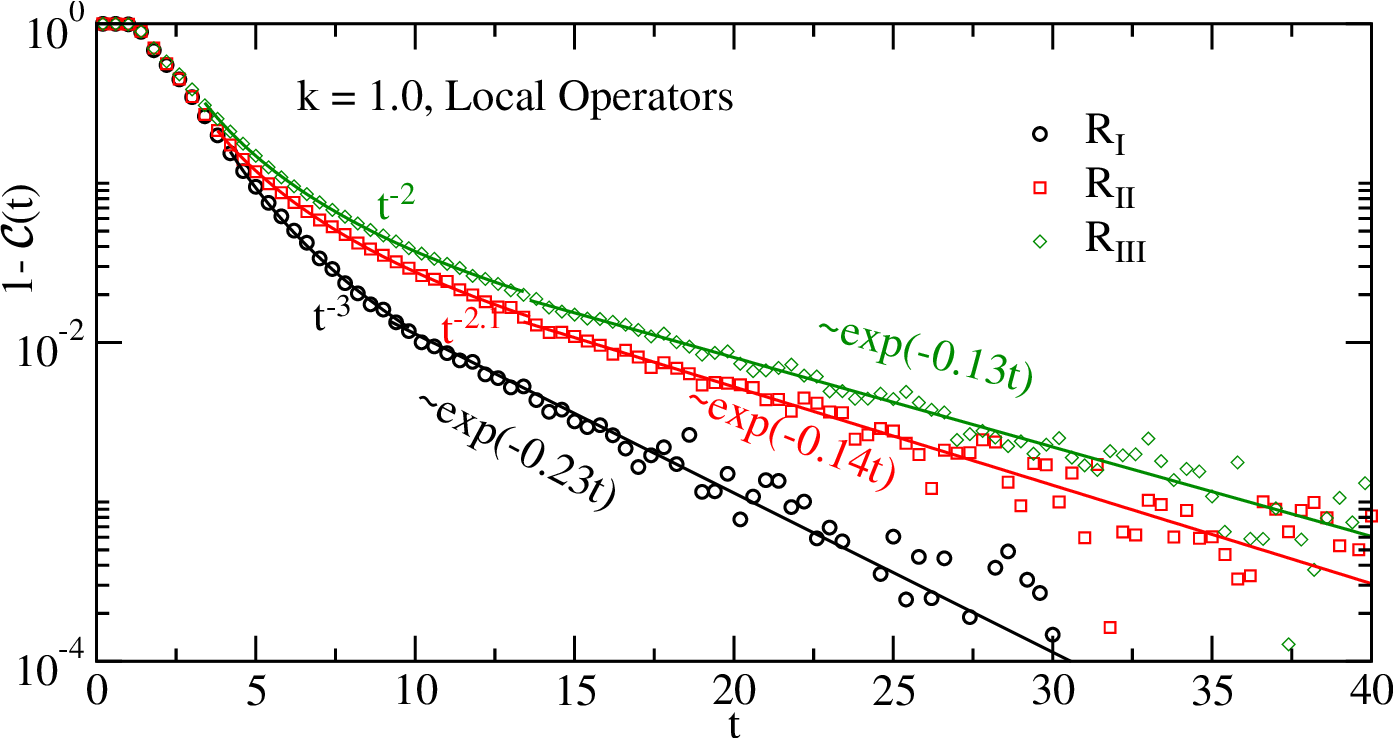}
	\includegraphics[width=\linewidth]{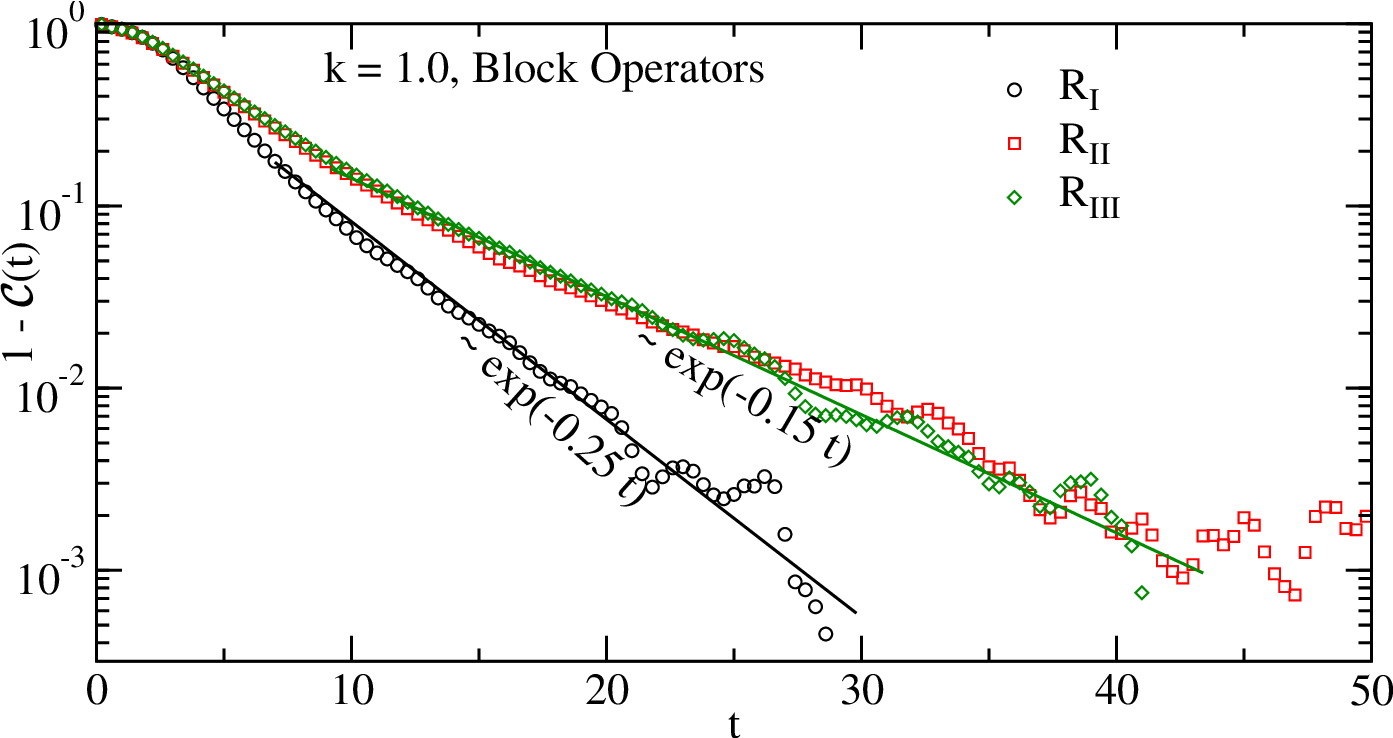}
\caption{Relaxation dynamics of OTOC for  field strengths $k = 1.0$ with local (top) as well as block (bottom) operators.}
\label{fig-otoc-relax-local-nonlocal-k1.0}
\end{figure}
Plots in Figs.~(\ref{fig-otoc-relax-local-nonlocal-k0.02}-\ref{fig-otoc-relax-local-nonlocal-k1.0}) illustrate normalized and inverted OTOCs, $(1-\mathcal C(t))$ for the Hamiltonian $H_\text{II}$. Several values of $k$ are considered to distinguish the dynamics during the crossover and also to analyze the effect of local, \textit{i.e.}, $A_1$ and $B_1$, and non-local  operators, $A_2$ and $B_2$.
As shown in Fig.~(\ref{fig-otoc-relax-local-nonlocal-k0.02}) for $k = 0.02$, the OTOC ($1-\mathcal C(t)$) decays as a power law. The NNSD and number variance statistics show nearly Poisson statistics for such small field strengths. For $k = 0.05$, the NNSD is still not wigner but show a linear repulsion and $\Sigma^2$ a logarithmic growth for smaller correlation lengths, but at larger values, both the quantities show nearly Poisson statistics. Consistent with these results, we observe the onset of exponential relaxation. The initial decay of $(1-\mathcal C(t))$ is power-law ($1/t$) followed by an exponential decay, as shown in Fig.~(\ref{fig-otoc-relax-local-nonlocal-k0.05}). However, the relaxation can still be assumed as a power-law because the exponent is very small ($O(10^{-3})$).
The chaotic regime corresponds to field strengths $0.5 \le k \le 1$ where both NNSD and number variance show an excellent agreement with RMT. For $k = 0.5$, as shown in Fig.~(\ref{fig-otoc-relax-local-nonlocal-k0.5}), the initial decay is power-law ($\approx 1/t$). Eventually, OTOC relaxes as an exponential after the transient power-law decay. We, however, observe a completely power-law decay with some of the realizations for block operators. We obtain similar results for $k = 1$, in Fig.~(\ref{fig-otoc-relax-local-nonlocal-k1.0}). The initial relaxation is still a power-law but with a high decay rate ($1/t^2$ to $1/t^3$ for different realizations).

\textit{Universality with respect to different realizations of random fields:} The OTOC exhibits quite distinct behavior in the scrambling and relaxation regimes. In the scrambling regime, for a fixed disorder strength, $k$, the growth rate is found to be the same for different realizations of random fields. But for the chaotic counterpart of the system ($0.5 \le k \le 1$), the rate show sensitive dependence to different realizations in the relaxation regime. The saturation regime also fluctuates with different realizations. In the next section, we develop an analytical framework to evaluate saturation through ETH.  The source of relaxation rate is still not quite clear, but we believe that eigen-state to eigen-state fluctuations and ETH can be useful in the evaluation of the relaxation rate. 

A similar characteristic of OTOC has also been observed in the OTOC for weakly coupled bipartite systems of kicked rotors in the context of intra-subsystem and inter-subsystem scrambling \cite{prakash-scrambling}. To summarize briefly, the system is chaotic when the coupling strength is not too weak. The OTOC grow exponentially upto the Ehrenfest time with a rate that does not depend on the coupling parameter, but when OTOC enters into the relaxation regime, the relaxation rate is found to be dependent only on the coupling. The relaxation rates and their dependence on the coupling have been evaluated by employing embedded random matrices for strongly chaotic subsystems. We believe that the strong sensitivity to details of the system in the relaxation regime is a key property of the relaxation regime and exist in all quantum systems.

\subsection{OTOC saturation and ETH}
\begin{figure}
	\includegraphics[width=\linewidth]{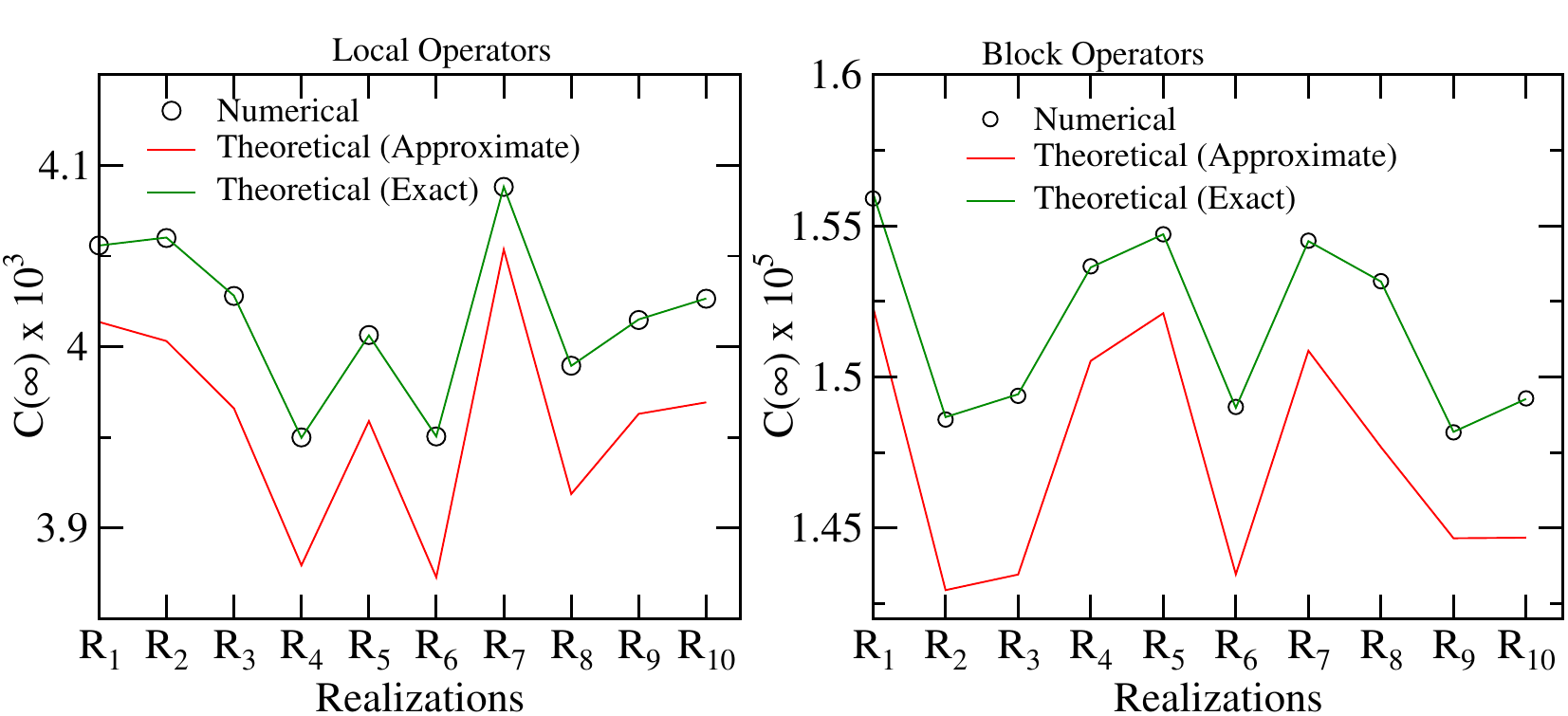}
\caption{The long-time saturation value of OTOC $C(\infty)$ is shown for different realizations of random fields in $H_{II}$ at disorder strength $k = 1.0$, for local operators ($A_1$ and $B_1$) and block operators ($A_2$ and $B_2$). The plots demonstrate an excellent agreement with Eq.~(\ref{eq-otoc-sat-formula-exact}). The approximate expression given by Eq.~(\ref{eq-otoc-sat-formula-approx}) also fluctuates in a qualitative similar manner for different realizations.}.
\label{fig-otoc-sat}
\end{figure}

The long-time saturation of OTOC is shown in Fig.~(\ref{fig-otoc-sat}) for the chaotic dynamics. The saturation value varies across different realizations of the random fields. To understand this behavior, we analyze the OTOC in the energy eigenbasis and, using ETH, obtain an estimate for the saturation value corresponding to different disorder realizations. For chaotic systems, RMT is also commonly used to estimate the OTOC saturation value \cite{prakash-scrambling, Balachandran-2021, PRB-block, lakshminarayan2019out, varikuti2024out}. This approach assumes that, at sufficiently long times, the evolution operator $e^{-iHt/\hbar}$ can be effectively modeled as a random unitary matrix. The resulting saturation value, obtained after averaging over random matrices, depends only on the choice of observables and the Hilbert space dimension. In the present case, however, the RMT prediction deviates significantly from the observed behavior, with the saturation value exhibiting a clear dependence on both the Hamiltonian and the observables.

\begin{figure}
	\includegraphics[width=0.49\linewidth]{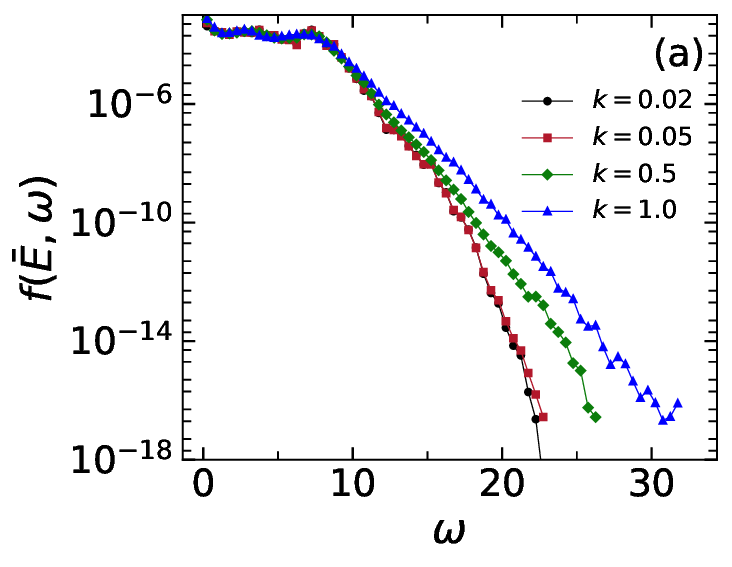}
		\includegraphics[width=0.49\linewidth]{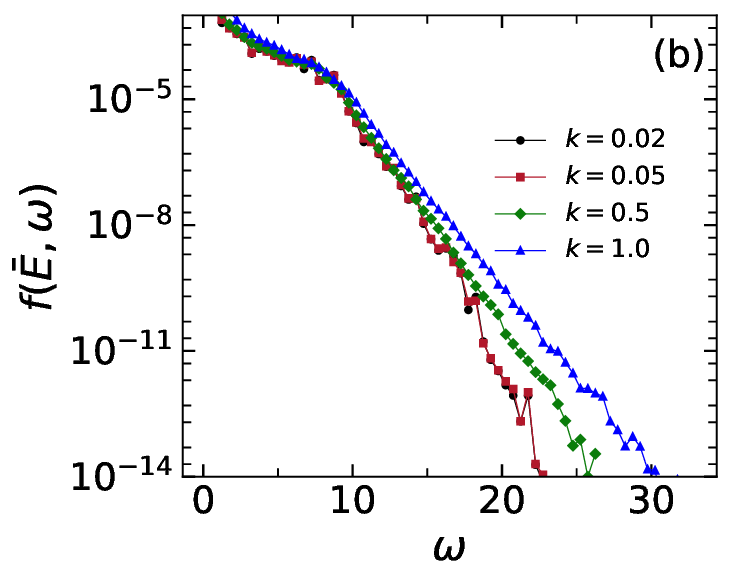}
\caption{The function $f(\bar{E}, \omega)$ is plotted as a function of $\omega$ for (a) local operators and (b) block operators. The results are shown for a single realization at different values of the disorder strength $k$. }
\label{fig-fvsw}
\end{figure}

According to ETH, any physical observable in the energy eigenbasis can be written as,
\begin{align}
	O_{mn} = {O}(\bar E)\delta_{mn} + e^{-S(\bar E)/2} f(\bar E, \omega) R_{mn},
	\label{eth}
\end{align}
where $\bar E=(E_m+E_n)/2$ and $\omega=E_m-E_n$. Here $S(\bar E)$ is the thermodynamic entropy, $O(\bar E)$ and $f(\bar E,\omega)$ are smooth functions of their arguments, and the $R_{mn}$ are random variables with zero mean and unit variance \cite{rigol2008thermalization,srednicki-eth}. The integrable systems appear almost diagonal in the energy-eigenbasis, and the function $f(\bar E, \omega)$ decays very rapidly with $\omega$. On the other hand, for a random matrix, \textit{e.g.}, GOE, $f(\bar E, \omega) \approx $ constant. For the Hamiltonian $H_\text{II}$, the function $f(\bar E, \omega)$ remains constant up to a critical $\omega_c$ and then decays exponentially with a rate which decreases for increasing disorder strength as shown in Fig.~(\ref{fig-fvsw}). The function $f(\bar E, \omega)$ remains the same for different realizations with the same disorder strength.

The fluctuations in saturation value in Fig.~(\ref{fig-otoc-sat}) indicate that the eigenstate-to-eigenstate fluctuations are not similar across the ensemble. To get an analytical expression, the OTOC in the energy eigenbasis can be written as $C(t) = C_2(t) - C_4(t)$, with
\begin{align}
	\nonumber   C_4(t) & = \sum_{j,k,l,m} e^{-i(E_j-E_k+E_l-E_m)t} \tilde{A}_{jk}\tilde{B}_{kl}\tilde{A}_{lm}\tilde{B}_{mj}\\
	\text{and}~C_2(t) & = \sum_{j,k} e^{i(E_j-E_k)t} (\tilde A^2)_{jk}(\tilde B^2)_{kj},
\end{align}
where $\tilde A$ represents the observable $A$ in the energy eigenbasis. The saturation at long times consists of terms with $E_j+E_l-E_k-E_m = 0$ and $E_j-E_k = 0$ in the expressions of $C_4$ and $C_2$, respectively. Generic chaotic systems due to eigen-energy repulsions are strongly non-degenerate. The first condition for chaotic systems can be written as $E_j = E_k$ and $E_l = E_m$, or $E_j = E_m$ and $E_l = E_k$ in the expression for $C_4$ \cite{Huang-PRL-2019}. The saturation value is therefore given by,
\begin{align}
	\label{eq-otoc-sat-formula}
\nonumber	C(\infty)& = \sum_j (\tilde A^2)_{jj} (\tilde B^2)_{jj} -(\tilde A_{jj})^2 (\tilde B_{jj})^2 \\
	& - \sum_{j,k\ne j} \left[\tilde A_{jj}\tilde A_{kk} |\tilde B_{jk}|^2 + \tilde B_{jj}\tilde B_{kk} |\tilde A_{jk}|^2\right] \\
	\nonumber \approx & \sum_j (\tilde A^2)_{jj} (\tilde B^2)_{jj} -(\tilde A_{jj})^2 (\tilde B_{jj})^2 \\
	\label{eq-otoc-sat-formula-exact}
	& - \hspace{-0.9cm}\sum_{j,k:|E_j-E_k|\approx \text{small}} \hspace{-0.9cm} \left[\tilde A_{jj}\tilde A_{jj} |\tilde B_{jk}|^2 + \tilde B_{jj}\tilde B_{jj} |\tilde A_{jk}|^2\right].
\end{align}
where in the last step, we have limited the summation over $k$.  The restricted summation ignores the off-diagonal terms for large $\omega$ because the function $f(\bar E,\omega)$  decays exponentially. We have also replaced $\tilde A_{kk}$ and $\tilde B_{kk}$ by $\tilde A_{jj}$ and $\tilde B_{jj}$ respectively, assuming that the observables $A$ and $B$ change smoothly and have almost similar values for nearby energies. The simplified saturation value is then given by
\begin{align}
	\label{eq-otoc-sat-formula-approx}
	C(\infty) = \sum_j \left[(\tilde A^2)_{jj} - (\tilde A_{jj})^2 \right] \left[(\tilde B^2)_{jj} - (\tilde B_{jj})^2 \right]
\end{align}

Eq.~(\ref{eq-otoc-sat-formula-exact}) exhibits an excellent agreement with numerical results as shown in Fig.~(\ref{fig-otoc-sat}). The simplified form given by Eq.~(\ref{eq-otoc-sat-formula-approx}) also show a fairly good agreement. Importantly, it preserves the fluctuations in saturation values for different realizations.

\begin{figure}[!ht]
	\centering
	\includegraphics[width=\linewidth]{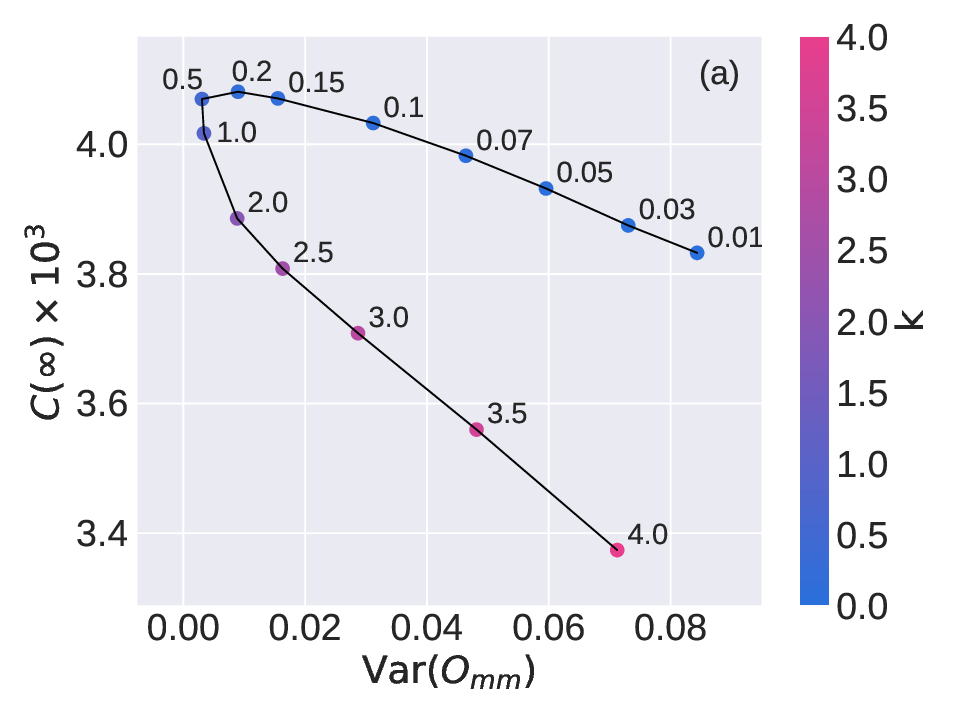}
	\includegraphics[width=\linewidth]{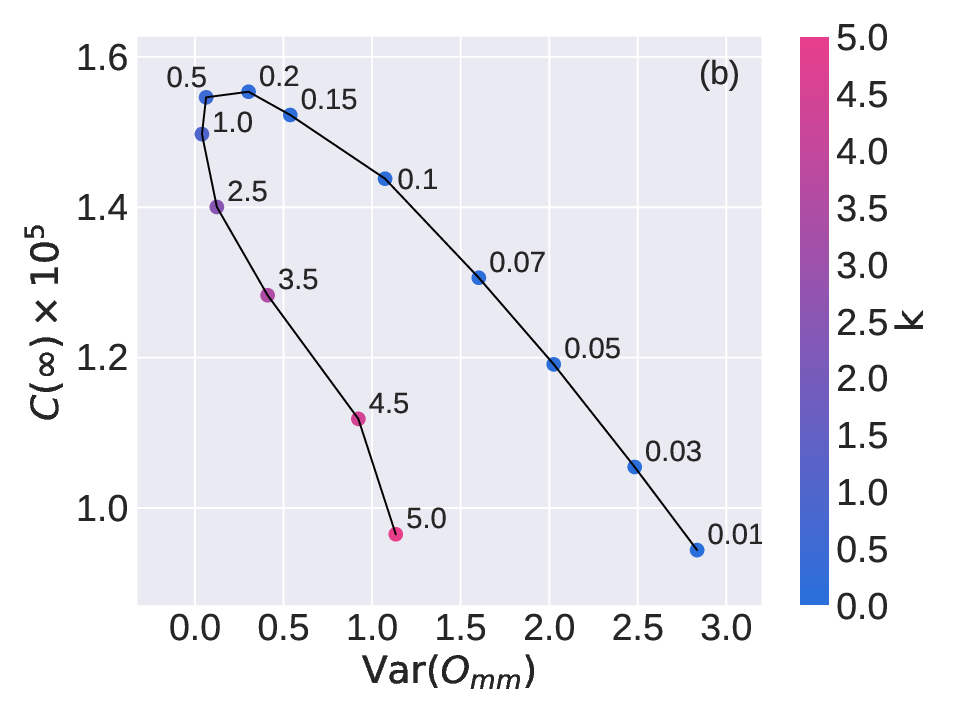}
	
	%\hspace{5mm}
	\caption{Long-time saturation value of the OTOC, $C(\infty)$,
	plotted as a function of the variance of the diagonal elements of the (a) Local operator $B_{1}$ and (b) Block operator $B_{2}$, in the eigenbasis of $H_\text{II}$ for disorder strengths in the range $0.01 < k < 5.0$. }	
	\label{fig-var-sat-spectrum}
\end{figure}

We now examine how the long-time OTOC saturation value and the matrix elements in the energy basis are modified with disorder strength. As the chaos increases in the system, the elements corresponding to higher $\omega$ also contribute to the dynamics, resulting in a higher saturation value and slower exponential decay of the function $f(\bar E, \omega)$. The variance of off-diagonal elements ($\overline{O_{mn}^2} = \frac{2}{N^2}\sum_{m,n\ne m} O_{mn}^2$) therefore increase or that of diagonal ($\overline{O_{mn}} = \frac{1}{N}\sum_{m} O_{mm}^2$) decreases.

In Fig.~(\ref{fig-var-sat-spectrum}), we show the variation of OTOC saturation values averaged over several realizations with diagonal variance for several values of disorder parameter $k$ in the range $0.01 \le k \le 5$. We observe that the variance decreases and saturation value rises monotonically with increasing $k$, indicating delocalization of the operator in the energy eigenbasis. The system is maximally thermalized for $0.2 \le k \le 1$. This is the same region where the system follows chaos properties fairly well (Fig.~(\ref{fig-summary})). 
At higher disorder strengths ($k \ge 1$), the thermalization is suppressed due to MBL effects. The diagonal variance increases and accordingly, the  OTOC exhibit a drop in its saturation value.

\section{Summary} \label{sec-summary}

In this paper, we investigate the  OTOC for the random-field Heisenberg XXZ spin chain to study chaos imprints in different temporal regimes. For an ensemble of such spin systems consisting of different realizations of random on-site fields, we analyze the spectral properties of the Hamiltonian across different disorder strengths and observe crossovers as the system transitions from the integrable regime to the chaotic regime, and eventually to the MBL phase. The short-range and long-range spectral fluctuations are characterized using the nearest neighbor spacing distribution and the number variance, respectively.

Our studies on OTOC focus on  three distinct dynamical regimes: scrambling, relaxation, and saturation. We systematically examine the behavior of the OTOC in each regime for both local and non-local observables. We find that the OTOC exhibits a power-law growth for both chaotic and integrable systems in the scrambling regime. Furthermore, the growth rate of the OTOC remains the same for the different realizations of the Hamiltonian.

We show that the relaxation regime of the OTOC can distinguish between integrable and chaotic dynamics.
We find that the relaxation rates are approximately the same for both local and non-local observables, but depend sensitively on the different realizations of the random field. For a weak disorder, where the system displays integrable spectral statistics, the OTOC exhibits a power-law ($1/t$) decay in the relaxation regime. With increasing disorder strength, the relaxation takes transitions to a power law of higher order followed by an exponential relaxation, marking the onset of chaos. We also study the saturation regime of the OTOC under the framework of the ETH. The saturation values also differ for different realizations of random fields, and are successfully explained using ETH. The saturation values (averaged over different realizations) show a qualitative inverse relationship with the variance of diagonal entries of one of the OTOC observable, written in energy epigenesis, and increases as systems approaches strong non-integrability, becomes maximum when spectra are RMT-like, and thereafter starts decreasing due to MBL effects. We believe that the relaxation regime can also be dealt with analytically using ETH.

\begin{acknowledgments}
The authors acknowledge Arul Lakshminarayan for a useful discussion on the present work and his suggestions to improve the manuscript.
\end{acknowledgments}

	\bibliographystyle{apsrev4-2}
	\bibliography{citation}  % Create a references.bib file
	
\end{document}